\documentclass{article}
\usepackage[utf8]{inputenc}
\usepackage[T1]{fontenc}
\usepackage[a4paper, hmargin = 2cm, bmargin = 2cm, tmargin = 2cm]{geometry}
\usepackage[tbtags]{amsmath}
\usepackage{amssymb}
\usepackage{pxfonts}
\usepackage[french,english]{babel}
\usepackage{ragged2e}
\usepackage{array}
\newcolumntype{x}[1]{>{\RaggedRight\noindent}p{#1}}
\usepackage{graphicx}
\usepackage{xspace}
\usepackage{maths_MF}
\usepackage{icomma_MF}
\usepackage{sections_MF}
\usepackage{appendix_MF}
\usepackage{textcomp}
\usepackage{notes_MF}
\usepackage{listes_MF}

\makeatletter
\renewcommand{\item@sep}{0.5ex}
\makeatother
\usepackage{natbib}
\bibpunct{(}{)}{;}{a}{,}{,~}
\usepackage{journal_abbrev}
\usepackage{indentfirst}
\newcommand\Info[1]{\text{\guilsinglleft}\texttt{#1}\text{\guilsinglright}}
\newcommand\Fichier[1]{\text{\guillemotleft}\texttt{#1}\text{\guillemotright}}
\newcommand{\Aspects}{\textsc{Aspects}\xspace}
\newcommand\pangle[2]{\angle(#1, #2)^+}
\newcommand\mangle[2]{\angle(#1, #2)^-}
\newcommand\oto{{\text{o:o}}}
\newcommand\sto{{\text{s:o}}}
\newcommand\ots{{\text{o:s}}}
\newcommand{\Prob}{P}
\newcommand{\Poto}{\Prob_{\!\oto}}
\newcommand{\Psto}{\Prob_{\!\sto}}
\newcommand{\Pots}{\Prob_{\!\ots}}
\newcommand\Lh{L}
\newcommand{\Lhoto}{\Lh_\oto}
\newcommand{\Lhsto}{\Lh_\sto}
\newcommand{\Lhots}{\Lh_\ots}
\renewcommand\card{\mathop{\diese}\mathopen{}}
\newcommand\paper{P-}
\newcommand\cf{cf{.}\xspace}
\newcommand\ie{i.e{.}\xspace}
\newcommand\eg{e.g{.}\xspace}
\newcommand\resp{resp{.}\xspace}
\newcommand\souligne[1]{\textbf{\textsc{#1}}}
\newcommand\pj{'_{\smash[t]{j}}}
\newcommand\pk{'_{\smash[t]{k}}}
\usepackage{typo_MF}
\newlength\bigrowspace
\newlength\smallrowspace
\usepackage{hyperref}
\begin{document}
\belowdisplayskip=\belowdisplayshortskip
\abovedisplayskip=\belowdisplayshortskip
\setlength{\bibhang}{\parindent}
\begin{center}
  \thispagestyle{empty}
  \leavevmode
  \vfill
  {\huge\bfseries The \Aspects code\\ 
    for probabilistic cross-identification\\ 
    of astrophysical sources\\[\baselineskip]}
  {\Large 
    %Version 2.0.1, 2014-4-29%
    Version 2.0.2, 2014-10-22%
    \\[\baselineskip]
    \Fichier{\href{http://www2.iap.fr/users/fioc/Aspects/}{www2.iap.fr/users/fioc/Aspects/}}
  }
  \bigskip
  \vfill
  \includegraphics[width=0.5\textwidth]{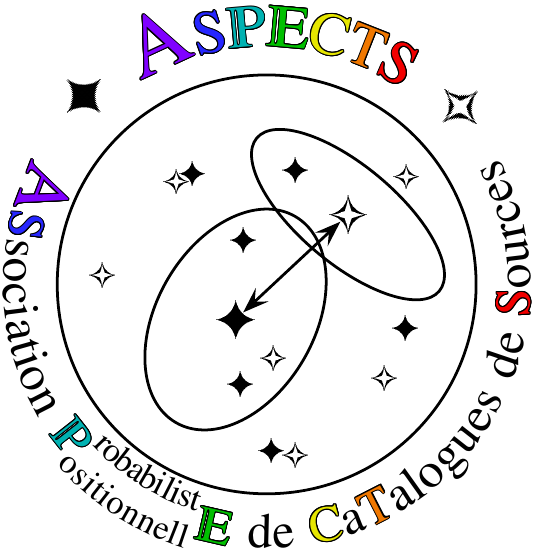}
  \vfill
  \bigskip
  {\LARGE\bfseries
    Documentation and complements\\[\baselineskip]}
  {\Large Michel \textsc{Fioc}\\[\baselineskip]}
  {\large
    \Info{Michel.Fioc@iap.fr}\\[\baselineskip]
    Institut d'astrophysique de Paris, \\
    université Pierre et Marie Curie, \\
    France}
  \vfill
\end{center}
\newpage
\tableofcontents
\newpage
\centerline{\bfseries\LARGE Documentation}
\addtocontents{toc}{\bigskip\bigskip\centerline{\bfseries\large Documentation}\medskip}
\vspace*{1cm}
\section{Introduction}
\label{intro}
\Aspects%
\footnote{%
  Pronounced [asp$\varepsilonup$] in
  International Phonetic Alphabet transcription,
  not [{\ae}spekts], and
  preferentially written ``\!Aspects'' if small capitals are not available.
  ``\!\Aspects'' is a French acronym for
  ``\!\emph{\souligne{As}sociation
    \souligne{p}ositionnell\souligne{e}/%
    \souligne{p}robabilist\souligne{e}
    de \souligne{c}a\souligne{t}alogues de
    \souligne{s}ources}''
  (``Positional/\allowbreak proba\-bilistic association of catalogs of
  sources'' in English).
  The French word ``\!\emph{Aspect}'' (pluralized in the acronym)
  has the same general meaning as the corresponding English word;
  interestingly, it signifies in particular the relative position of heavenly
  bodies\textellipsis{}
}
is a code for the probabilistic cross-identification of astrophysical sources.
Version~2.0\footnote{%
  The current version is in fact v2.0.2. 
  The only differences with respect to v2.0.1 are in the documentation 
  (introduction of hyperlinks, exact reference for the main paper).
  Version v2.0.1 differed from v2.0[.0] by a few
  minor changes in the formatting of outputs produced by 
  \Fichier{example\_read.f90} and by 
  subroutine \Info{write\_prob} of \Fichier{mod\_output\_prob.f90}.
}
 of the code fully supersedes the version~1.0 described in
\citet{article_Aspects_v1}%
\footnote{The latter provided biased probabilities (and therefore also
  biased values of the likelihood and of estimators of unknown parameters)
  under assumption $H_\oto$ (\cf~\S~\ref{purpose}).}
and is an implementation in Fortran~95
of the relations established in \citet{article_Aspects_v2}%
\footnote{Available as file \Fichier{paper.pdf} in directory \Fichier{v2.0/}
  (see below). References to sections or equations of this paper are preceded by ``\paper''.
  A basic presentation of \Aspects is also given in
  \citet{seminaire_Aspects} (file \Fichier{IAP\_seminar.pdf}).}.
Its source files are freely%
\footnote{The \emph{Numerical Recipes} routines in Fortran~90 \citep{NR}
  mentionned in \citet{article_Aspects_v2} have been replaced by free
  equivalents.}
available at
\begin{center}
  \Fichier{\href{http://www2.iap.fr/users/fioc/Aspects/}{http://www2.iap.fr/users/fioc/Aspects/}}
\end{center}
in tar file \Fichier{Aspects\_v2.0.tar}.
Type \Info{tar xvf Aspects\_v2.0.tar} to extract them in
\Fichier{Aspects\_v2.0/}.
\section{Overview}
\label{purpose}
Given two catalogs $K$ and $K'$ of $n$ and $n'$ astrophysical sources,
respectively,
\Aspects computes, for any objects $M_i \in K$
and $M\pj \in K'$, the probability that $M\pj$ is a
\emph{counterpart} of
$M_i$, \ie\ that they are the same source.
To determine this \emph{probability of association},
denoted by $P(A_{i, j} \mid C \cap C')$ in \citet{article_Aspects_v2},
the code
takes into account the coordinates and positional uncertainties
of all the $K$-sources ($C$) and of all the $K'$-sources ($C'$).
\Aspects also computes the probabilities
$P(A_{i, 0} \mid C \cap C')$ and $P(A_{0, j} \mid C \cap C')$ that
$M_i$ or $M\pj$ has no counterpart.

Three exclusive hypotheses are considered for these calculations:
\begin{itemize}
\item{\bfseries\mathversion{bold}$H_\sto$}\,:
  a $K$-source has at most one counterpart in $K'$,
  but a $K'$-source may have several counterparts in~$K$\linebreak
  ({\bfseries\mathversion{bold}several-to-one $K$-$K'$ associations}).
\item{\bfseries\mathversion{bold}$H_\ots$}\,:
  a $K'$-source has at most one counterpart in $K$,
  but a $K$-source may have several counterparts in $K'$\linebreak
  ({\bfseries\mathversion{bold}one-to-several $K$-$K'$ associations}).
\item{\bfseries\mathversion{bold}$H_\oto$}\,:
  a $K$-source has at most one counterpart in $K'$
  and vice versa ({\bfseries\mathversion{bold}one-to-one associations}).
\end{itemize}
All quantities derived under either $H_\sto$, $H_\ots$ or $H_\oto$
have a subscript ``\sto'', ``\ots'' or ``\oto'', respectively.

Although $H_\oto$ is the most natural assumption because of its symmetry,
the other two deserve some consideration too.
For instance, if the spatial resolution is much lower in $K'$ than in $K$,
several distinct $K$-sources may be merged in $K'$; $H_\sto$
might then be preferred.
On the opposite, if objects are more patchy at the wavelength of $K'$ than at
that of $K$, single extended $K$-sources may look as several
objects in $K'$; $H_\ots$ should be more appropriate in this case.
Additional reasons for making computations under $H_\sto$ and $H_\ots$
is that they are much simpler and faster than under $H_\oto$,
and that they are used in the first step of the iterative procedure
leading to the probabilities $\Poto(A_{i, j} \mid C \cap C')$\textellipsis

To compute the probabilities of association, \Aspects needs the
a~priori (\ie, ignoring $C$ and $C'$) probability $f$ (or $f'$)
that any object in $K$ (or $K'$) has a counterpart in the other catalog.
The code obtains estimates $\hat f_\sto$, $\hat f'_\ots$ and $\hat f_\oto$
(or $\hat f'_\oto = n*\hat f_\oto/n'$) of $f$ or $f'$ by maximizing
the \emph{likelihood} $L$ to observe all the sources at their effective
positions under assumptions $H_\sto$, $H_\ots$ and $H_\oto$, respectively.

\Aspects computes these likelihoods, and in particular their maxima
$\hat\Lh_\sto \egdef \Lh_\sto(\hat f_\sto)$,
$\hat\Lh_\ots \egdef \Lh_\ots(\hat f'_\ots)$
and $\hat\Lh_\oto \egdef \Lh_\oto(\hat f_\oto)$%
\footnote{All quantities computed at $f = \hat f$ or $f' = \hat f'$
  under the corresponding assumption bear a circumflex accent hereafter.
  For instance, $\expandafter\hat\Psto(A_{i, j} \mid C \cap C')$ is the value
  of $\Psto(A_{i, j} \mid C \cap C')$ for $f = \hat f_\sto$.}.
The assumption for which the maximum likelihood
is the largest should be the most appropriate to compute the probabilities
of association.
\nopagebreak
\section{Contents of the \Fichier{Aspects\_v2.0/}  directory}
\label{contents}
A large variety of cases may be considered, depending on whether the positional
uncertainties are known in both catalogs, only one or none.
If not fully known, they may be modelled with some additional
parameters (\eg\ $\mathring\sigma$ and $\mathring\nu$;
cf.~\S~\paper A),
to be determined by likelihood maximization;
galaxy positional uncertainties, for instance, might be expressed as a
function of the size of objects.

It would be impractical to cover all the cases in a single
general-purpose main program. Therefore, besides
the main module, \Fichier{mod\_Aspects.f90}, and a series of minor
modules, we provide two main programs as examples:
\Fichier{example\_simul.f90} and \Fichier{example\_read.f90}
(cf.~\S~\ref{run}).
These programs assume that positional uncertainties are known,
but extensive comments should make them easy to adapt to other cases
(cf.~\S~\ref{adapt}).

The complete list of files in \Fichier{Aspects\_v2.0/}
is the following:
\begin{center}
  \begin{tabular}{lll}
    \hline
    \Fichier{Read\_me.pdf} &
    \Fichier{paper.pdf} &
    \Fichier{Makefile} \\[\bigrowspace]
    \Fichier{example\_simul.f90} &
    \Fichier{example\_read.f90} &
    \Fichier{mod\_Aspects.f90} \\[\bigrowspace]
    \Fichier{mod\_simul\_catalogs.f90} &
    \Fichier{mod\_output\_prob.f90} &
    \Fichier{mod\_read\_catalog.f90} \\[\bigrowspace]
    \Fichier{mod\_variables.f90} &
    \Fichier{mod\_constants.f90} &
    \Fichier{mod\_types.f90} \\[\bigrowspace]
    \Fichier{mod\_files.f90} &
    \Fichier{mod\_heap\_index.f90} &
    \Fichier{input\_example\_simul.dat} \\[\bigrowspace]
    \Fichier{input\_example\_read.dat} &
    \Fichier{K\_p.dat} &
    \Fichier{K\_u.dat} \\[\bigrowspace]
    \Fichier{IAP\_seminar.pdf} & & \\
    \hline
  \end{tabular}
\end{center}
%
% The current version, v2.0.1, only differs from v2.0[.0] by a few
% minor changes in the formatting of outputs produced by 
% \Fichier{example\_read.f90} and by 
% subroutine \Info{write\_prob} of \Fichier{mod\_output\_prob.f90}.
% 
\section{Compilation}
\label{compil}
The default compiler is \Info{ifort}. To choose instead \Info{gfortran},
edit file \Fichier{Makefile}: comment line~$4$ (insert a \Info{\diese}
before \Info{f95\_compiler}) and uncomment line~$9$.
The code is written in standard Fortran~95, so other compilers should also work.
Note that none of the options of compilation (set by variable
\Info{f95\_options}) is mandatory: they provide various checks during
the compilation and the execution, but might actually slow down the latter.

To compile the code, just type \Info{make} on the keyboard.
This will produce two executables: \Fichier{example\_\allowbreak simul} and
\Fichier{example\_read}. To recompile from scratch, type first
\Info{make clean\_all}: this will delete the executables and the
objects (\Fichier{{\itshape*}.o}) and \Fichier{{\itshape*}.mod} files.
To clean the directory while keeping the executables, type just
\Info{make clean}.
\section{Execution}
\label{run}
\subsection{Code \Fichier{example\_simul}}
This executable simulates a sequence of twin mock catalogs and analyzes them.
\subsubsection{Inputs}
\Fichier{example\_simul} first asks for the following input parameters:
\begin{center}
  \begin{tabular}{lll}
    \hline
    \Info{n\_simul} & integer & Number of simulations requested. 
    \\
    \hline
    \Info{n\_u} & integer & Number $n$ of $K$-sources. 
    \\[\smallrowspace]
    \Info{n\_p} & integer & Number $n'$ of $K'$-sources. 
    \\
    \hline
    \Info{input\_f\_u} & real & Requested fraction $f$ of $K$-sources with a
    counterpart in $K'$. 
    \\
    \hline
    \Info{semi\_axis\_a\_u} & real & $1$-$\sigma$ positional uncertainty 
    in radians along the major axis
    \\
    &&\quad for $K$-sources. 
    \\[\smallrowspace]
    \Info{semi\_axis\_b\_u} & real & The same along the minor axis. 
    \\[\smallrowspace]
    \Info{semi\_axis\_a\_p} & real & $1$-$\sigma$ positional uncertainty 
    in radians along the major axis 
    \\
    &&\quad for $K'$-sources. 
    \\[\smallrowspace]
    \Info{semi\_axis\_b\_p} & real & The same along the minor axis.
    \\
    \hline
    \Info{one\_to\_one} & boolean & \Info{T} for simulations of 
    one-to-one $K$-$K'$ associations; 
    \\
    & & \quad \Info{F} for simulations of several-to-one 
    $K$-$K'$ associations. 
    \\
    \hline
    \Info{area\_S} & real & Area $S$ in steradians of the surface covered by
    $K$ and $K'$; 
    \\
    && \quad set to $4*\piup$ if the input value is not in $\IOF]0, 4*\piup]$.
    \\
    \hline
    \Info{output\_catalogs} & boolean & \Info{T} to write the catalogs 
    created by the \emph{first} simulation; 
    \\
    & & \quad \Info{F} otherwise. 
    \\[\smallrowspace]
    \Info{K\_u\_file} &
    string of &
    Name of the output file for catalog $K$; 
    \\
    &\quad characters& \quad not required if \Info{output\_catalogs} is false. 
    \\[\smallrowspace]
    \Info{K\_p\_file} & string & The same for catalog $K'$.
    \\
    \hline
  \end{tabular}
\end{center}

A set of input parameters is given as an example in
\Fichier{input\_example\_simul.dat}.
Type
\[
\Info{example\_simul < input\_example\_simul.dat}
\]
to run the code with these values.
\subsubsection{Outputs}
\label{example_simul_outputs}
If requested, \Fichier{example\_simul} writes the catalogs created by the
first simulation in the files designated by \Info{K\_u\_file} and
\Info{K\_p\_file}.
If any of these files already exists, new files with the same base names
and a suffix derived from the current date and time are created.

For each simulation, \Fichier{example\_simul} returns the following quantities:
\begin{center}
  \begin{tabular}{ll}
    \hline
    \multicolumn{2}{c}{\bfseries Properties of the simulation}
    \\[\smallrowspace]
    \Info{seed} & Array of integers generating the sequence of
    random numbers used 
    \\
    & \quad in the simulation. 
    \\[\smallrowspace]
    \Info{n\_unavailable} & Number of $K$-sources unassociated because no
    $K'$-source was available 
    \\
    & \quad ($\neq 0$ only if \Info{one\_to\_one} is true).
    \\[\smallrowspace]
    \Info{n\_side\_effects} & Number of $K$-sources unassociated because
    of side effects (only if $\Info{area\_S} < 4*\piup$).
    \\[\smallrowspace]
    \Info{eff\_f\_u} & Effective fraction of $K$-sources with a counterpart
    in the other catalog. 
    \\
    \Info{eff\_f\_p} & The same for $K'$-sources.
    \\[\smallrowspace]
    \Info{oto\_deviation} &
    $= n\times\text{\Info{eff\_f\_u}}/(n'\times\Info{eff\_f\_p})$.
    Deviation with respect to one-to-one associations 
    \\
    & \quad ($=1$ if \Info{one\_to\_one} is true;
    $>1$ otherwise, \ie\ for several-to-one associations).
    \\
    \hline
    \multicolumn{2}{c}{\bfseries Several-to-one computations}
    \\[\smallrowspace]
    \Info{sto\_f\_u} & $= \hat f_\sto$: maximum likelihood estimator
    of $f$ under assumption $H_\sto$. 
    \\[\smallrowspace]
    \Info{std\_dev\_sto\_f\_u} & Standard deviation of $\hat f_\sto$. 
    \\[\smallrowspace]
    \Info{sto\_f\_p} & $= \hat f'_\sto$: estimator of $f'$ under assumption
    $H_\sto$. 
    \\[\smallrowspace]
    \Info{max\_ln\_sto\_Lh} & $= \ln\hat\Lh_\sto$: maximum value of the
    log-likelihood under $H_\sto$.
    \\
    \hline
    \multicolumn{2}{c}{\bfseries One-to-several computations}
    \\[\smallrowspace]
    \Info{ots\_f\_p} &  $= \hat f'_\ots$: maximum likelihood estimator
    of $f'$ under assumption $H_\ots$. 
    \\[\smallrowspace]
    \Info{std\_dev\_ots\_f\_p} & Standard deviation of $\hat f'_\ots$. 
    \\[\smallrowspace]
    \Info{ots\_f\_u} & $= \hat f_\ots$: estimator of $f$ under assumption
    $H_\ots$. 
    \\[\smallrowspace]
    \Info{max\_ln\_ots\_Lh} & $= \ln\hat\Lh_\ots$:
    maximum value of the log-likelihood under $H_\ots$. 
    \\
    \hline
    \multicolumn{2}{c}{\bfseries One-to-one computations}
    \\[\smallrowspace]
    \Info{oto\_f\_u} & $= \hat f_\oto$: maximum likelihood estimator
    of $f$ under assumption $H_\oto$. 
    \\[\smallrowspace]
    \Info{std\_dev\_oto\_f\_u} & Standard deviation of $\hat f_\oto$. 
    \\[\smallrowspace]
    \Info{oto\_f\_p} & $= \hat f'_\oto = n*\hat f_\oto/n'$. 
    \\[\smallrowspace]
    \Info{max\_ln\_oto\_Lh} & $= \ln\hat\Lh_\oto$: maximum value of the
    log-likelihood under $H_\oto$. 
    \\
    \hline
  \end{tabular}
\end{center}
\subsection{Code \Fichier{example\_read}}
This executable reads two catalogs, analyzes them and computes the probabilities
of association of the sources they contain.
It also makes a few checks.
\subsubsection{Inputs}
\Fichier{example\_read} asks for the following inputs:
\begin{center}
  \begin{tabular}{ll}
    \hline
    \Info{K\_u\_file} & Name of the input file containing the coordinates
    and positional uncertainties
    \\
    & \quad  of $K$-sources.
    \\[\smallrowspace]
    \Info{K\_u\_file} & The same for $K'$-sources.
    \\
    \hline
    \Info{sto\_prob\_u\_file} & \vrule height 2.4ex depth 0pt width 0pt
    Name of the output file containing the probabilities 
    $\expandafter\hat\Psto(A_{i, j} \mid C \cap C')$ for all $M_i \in K$.
    \\[\smallrowspace]
    \Info{sto\_prob\_p\_file} & \vrule height 0pt depth 1.2ex width 0pt
    The same for all $M\pj \in K'$.
    \\
    \hline
    \Info{ots\_prob\_u\_file} & \vrule height 2.4ex depth 0pt width 0pt
    Name of the output file containing
    the probabilities $\expandafter\hat\Pots(A_{i, j} \mid C \cap C')$ 
    for all $M_i \in K$.
    \\[\smallrowspace]
    \Info{ots\_prob\_p\_file} & \vrule height 0pt depth 1.2ex width 0pt
    The same for all $M\pj \in K'$.
    \\
    \hline
    \Info{oto\_prob\_u\_file} & \vrule height 2.4ex depth 0pt width 0pt
    Name of the output file containing
    the probabilities $\expandafter\hat\Poto(A_{i, j} \mid C \cap C')$ 
    for all $M_i \in K$.
    \\[\smallrowspace]
    \Info{oto\_prob\_p\_file} & \vrule height 0pt depth 1.2ex width 0pt
    The same for all $M\pj \in K'$.
    \\
    \hline
    \Info{skip\_checks} & \Info{T} to skip checks of one-to-one computations;
    \Info{F} otherwise. 
    \\
    \hline
  \end{tabular}
\end{center}
(All the \Info{{\itshape*}\_file} inputs are strings of characters.)

A set of input parameters is given as an example in
\Fichier{input\_example\_read.dat}.
Type
\[
\Info{example\_read < input\_example\_read.dat}
\]
to run the code on the catalogs \Fichier{K\_u.dat} and \Fichier{K\_p.dat}
(these have been created by \Fichier{example\_simul}).
\Paragraph{Warning}
The subroutine \Info{read\_catalog} of
\Fichier{mod\_read\_catalog.f90} is called to read the files designated by
\Info{K\_u\_\allowbreak file} and \Info{K\_p\_file}
(the format of these files is described in the header of \Fichier{K\_u.dat}
and \Fichier{K\_p.dat}).
It reads in particular the surface area $S$ covered by
$K$ and $K'$.
This is used to set to $1/S$ the quantities $\xi_{i, 0}$ (\cf~\S~\paper3.1.1)
and $\xi_{0, j}$ (\cf~\S~\paper3.2.1)
for all $M_i$ and $M\pj$, assuming that unrelated sources
are uniformly and randomly distributed on the surface.
The user must provide this quantity for real catalogs.
If it is less than $4*\piup$~steradians, he also has to make sure that
the catalogs cover the same surface:
\Fichier{example\_read} checks only that the surface \emph{areas}
---not the \emph{surfaces} themselves--- are identical.
\subsubsection{Outputs}
\Fichier{example\_read} provides the following outputs:
\begin{center}
  \begin{tabular}{lll}
    \hline
    \multicolumn{3}{c}{\bfseries Fractions of sources with a counterpart}
    \\[\bigrowspace]
    \multicolumn{2}{l}{\Info{sto\_f\_u}, \Info{std\_dev\_sto\_f\_u}, 
      \Info{sto\_f\_p}} & Estimates
    under $H_\sto$ (\cf~\S~\ref{example_simul_outputs}).
    \\[\smallrowspace]
    \multicolumn{2}{l}{\Info{ots\_f\_u}, \Info{std\_dev\_ots\_f\_u}, 
      \Info{ots\_f\_p}} & The same under $H_\ots$. 
    \\[\smallrowspace]
    \multicolumn{2}{l}{\Info{oto\_f\_u}, \Info{std\_dev\_oto\_f\_u}, 
      \Info{oto\_f\_p}}& The same under $H_\oto$. 
    \\
    \hline
    \multicolumn{3}{c}{{\bfseries Choice of association model}%
      \footnotemark%
    } 
    \\[\bigrowspace]
    \multicolumn{2}{l}{\Info{max\_ln\_sto\_Lh}, \Info{max\_ln\_ots\_Lh}, 
      \Info{max\_ln\_oto\_Lh}} & Maximum values of the log-likelihood
    \\
    && \quad under $H_\sto$, $H_\ots$ and $H_\oto$
    (\cf~\S~\ref{example_simul_outputs}).
    \\
    \hline
    \multicolumn{2}{c}{\bfseries Probabilities of association} 
    \\[\bigrowspace]
    \Info{sto\_prob\_u($i$)} & Written in \Info{sto\_prob\_u\_file}. &
    Probabilities $\expandafter\hat\Psto(A_{i, j} \mid C \cap C')$ for all
    $M_i \in K$.
    \\[\smallrowspace]
    \Info{sto\_prob\_p($j$)} & Written in \Info{sto\_prob\_p\_file}. &
    The same for all $M\pj \in K'$.
    \\[\bigrowspace]
    \Info{ots\_prob\_u($i$)} & Written in \Info{ots\_prob\_u\_file}. &
    Probabilities $\expandafter\hat\Pots(A_{i, j} \mid C \cap C')$ for all
    $M_i \in K$.
    \\[\smallrowspace]
    \Info{ots\_prob\_p($j$)} & Written in \Info{ots\_prob\_p\_file}. &
    The same for all $M\pj \in K'$.
    \\[\bigrowspace]
    \Info{oto\_prob\_u($i$)} & Written in \Info{oto\_prob\_u\_file}. &
    Probabilities $\expandafter\hat\Poto(A_{i, j} \mid C \cap C')$ for all
    $M_i \in K$.
    \\[\smallrowspace]
    \Info{oto\_prob\_p($j$)} & Written in \Info{oto\_prob\_p\_file}. &
    \vrule height 0pt depth 1.2ex width 0pt
    The same for all $M\pj \in K'$.
    \\
    \hline
  \end{tabular}
  \footnotetext{\Fichier{example\_read} also recalls the recommendations
    given in the second paragraph of \S~\paper6.4.}
\end{center}
The quantities \Info{sto\_prob\_u($i$)}, etc., are records of type
\Info{prob\_struct} (see \Fichier{mod\_types.f90}).
Their fields are described in \S~\ref{prob_record}.
To read the list of possible counterparts
and the corresponding probabilities of association
from the output files designated by \Info{sto\_prob\_u\_file},
etc., see the header of these files.

As an illustration, \Fichier{example\_read} also computes $\ln\Lh_\oto(f)$
for a grid of equidistant values of $f$ with the subroutine
\Info{compute\_ln\_oto\_Lh} of \Fichier{mod\_Aspects.f90}%
\footnote{This subroutine has already been used to compute
  \Info{max\_ln\_oto\_Lh}. As calculations under $H_\oto$ are lengthy,
  \Info{compute\_ln\_oto\_Lh} may be called to compute $\ln\Lh_\oto$ for
  a given $f$ and for a grid of $f$ at the same time.}.
\subsubsection{Checks of results obtained under \texorpdfstring{$H_\oto$}{H\_\{o:o\}}}
\label{checks}
The code finally makes some checks if \Info{skip\_checks} is false.
It first computes $\ln\hat\Lh_\oto$ (\Info{max\_ln\_\allowbreak
  oto\_Lh\_\allowbreak alt}) again with the function \Info{func\_ln\_oto\_Lh\_alt}
of \Fichier{mod\_Aspects.f90},
which is based on the results of \S~\ref{other_Lh_oto}.
As currently implemented, this procedure is however extremely slow and should
generally be avoided.

\Fichier{example\_read} then swaps $K$ and $K'$, and computes
$\hat f_\oto$, $\hat f'_\oto$, $\ln\hat\Lh_\oto$ and the probabilities
$\expandafter\hat\Poto(A_{i, j} \mid C \cap C')$ again.
The suffix \Info{\_sw} is appended to all these quantities, as well as
to the files in which the probabilities are written.
\section{Understanding and adapting the code}
\label{adapt}
\subsection{Terminological and typographical conventions}
\label{notations}
Here is a brief summary of the conventions used in the code and this
file:
\makeatletter
\par\nobreak
\@afterheading
\makeatother
\begin{itemize}
\item
  In outputs to the terminal and comments, code fragments
  are written between \Info{\textasciigrave} (grave accents),
  and filenames between \Info{"}.
  Non Fortran mathematical expressions, quantities and objects
  defined in the article are between dollars in pseudo-\TeX.

  In the documentation and the complements,
  code fragments are surrounded by ``\guilsinglleft''
  and ``\guilsinglright''; file names, by ``\guillemotleft'' and
  ``\guillemotright'';
\item
  Quantities chiefly or only related to catalog $K$
  are written with a suffix \Info{\_u} (for ``unprimed'');
  those related to $K'$, with a suffix \Info{\_p} (for ``prime'').
  In file \Info{mod\_Aspects.f90}, \Info{\_1} and \Info{\_2} are used
  instead because $K$ and $K'$ are sometimes swapped;
\item
  The word ``neighbor'' (\Info{ngb}) designates a nearby object in the
  same catalog;
  ``counterpart'' (\Info{ctp}), a nearby object in the other
  catalog;
\item
  The notations \Info{sto}, \Info{ots} and \Info{oto} are used to distinguish
  the results obtained under the assumptions $H_\sto$, $H_\ots$ and $H_\oto$,
  respectively.
\end{itemize}
\subsection{Coordinates and positional uncertainties}
\begin{itemize}
\item
  The coordinates and positional uncertainties of $K$- and $K'$-sources
  are written in the records \Info{coord\_u} and \Info{coord\_p}.
  These records are of type \Info{coord\_struct} (see \Fichier{mod\_types.f90})
  and contain the following fields:
  \begin{center}
    \begin{tabular}{ll}
      \hline
      \Info{alpha} & Right ascension (or equivalent: Galactic longitude, etc.).
      \\[\smallrowspace]
      \Info{delta} & Declination (or Galactic latitude, etc.). 
      \\[\smallrowspace]
      \Info{semi\_axis\_a} & Semi-major axis of the positional uncertainty
      ellipse. 
      \\[\smallrowspace]
      \Info{semi\_axis\_b} & Semi-minor axis. 
      \\[\smallrowspace]
      \Info{beta} & Positional angle of the ellipse. 
      \\
      \hline
    \end{tabular}
  \end{center}
  For instance, the right ascension $\alpha_i$ of the $K$-source $M_i$
  is \Info{coord\_u($i$)\%alpha}, and the semi-minor axis $b\pj$ of the
  $K'$-source $M\pj$ is \Info{coord\_p($j$)\%semi\_axis\_b}.

  All these quantities are in radians.
\item
  Right ascensions are in $\IFO[0, 2*\piup[$. Sources must be ordered by
  increasing right ascension; this is done automatically when calling
  subroutine
  \Info{simul\_catalogs}, as in \Fichier{example\_simul}, but you have
  to do it yourself if you read catalogs, as in \Fichier{example\_read}
  (this is checked, however).
\item
  The positional uncertainty angle $\beta$ is counted eastward
  from the North to the semi-major axis and is in $\IFO[0, \piup[$.
\item
  The distribution of the observed position of a source relative to its true
  position is described by a $2$-dimensio\-nal normal (\ie\ Gaussian) law%
  \footnote{%
    The sphere is locally assimilated to its tangent plane or to a tangent
    cylinder. Note that this would not be appropriate if the semi-major axis
    were not negligible with respect to $\piup$~radians (!).
    A Kent distribution \citep{Kent}, for instance, would then
    seem a better choice,
    as it is one possible generalization of the normal law to the sphere,
    but, contrary to a Gaussian, it is not stable:
    the sum of independent Kent-distributed random variables does not
    follow a distribution of the same type.%
  }%
  .
  The semi-major (\resp\ semi-minor) axis is the $1$-$\sigma$ positional
  uncertainty along the major (\resp\ minor) axis, independently from the
  position along the minor (\resp\ major) axis. So, if $a$, $b$, $X$ and $Y$
  are, respectively, the semi-major and semi-minor axes and the positional
  errors along each axis,
  then
  \begin{equation}
    P\bigl((X, Y)\in \IFO[x, x+\df x[\times\IFO[y, y+\df y[\bigr) =
    \frac{1}{2*\piup*a*b}*\exp\left(-\frac{1}{2}*
      \left[\frac{x^2}{a^2}+\frac{y^2}{b^2}\right]\right)*\df x*\df y.
  \end{equation}
\end{itemize}
\subsection{Probabilities of association}
\label{prob_record}
For any source, the probabilities of association
are written in records of type \Info{prob\_struct}
(see \Fichier{mod\_types\allowbreak.f90})
containing the following fields:
\begin{center}
  \begin{tabular}{ll}
    \hline
    \Info{n\_ctp} & Number of possible real counterparts to the source.
    \\[\smallrowspace]
    \Info{ctp} & For $k \in \IE[1, \text{\Info{n\_ctp}}]$, \Info{ctp($k$)} is
    the ordinal number of the $k$-th counterpart; 
    \\
    & \quad the fictitious counterpart \Info{ctp(0)} ($= 0$) corresponds
    to no counterpart.
    \\[\smallrowspace]
    \Info{prob} & For $k \in \IE[1, \text{\Info{n\_ctp}}]$, \Info{prob($k$)}
    is the probability that the source is associated with \Info{ctp($k$)};
    \\
    & \quad \Info{prob(0)} is the probability that the source 
    has no counterpart.
    \\
    \hline
  \end{tabular}
\end{center}
For instance, \Info{sto\_prob\_u($i$)\%n\_ctp} is the number of possible
real counterparts of $M_i$, and \Info{sto\_prob\_u($i$)\%\allowbreak ctp($k$)}
is the ordinal number~$j$ of the $k$-th counterpart $M\pj$ of $M_i$ under
$H_\sto$.
Similarly, if $i = {}$\Info{oto\_prob\_p($j$)\%\allowbreak ctp($k$)},
\Info{oto\_\allowbreak prob\_p($j$)\%prob($k$)} is the probability
that $M\pj$ is associated to $M_i$ under $H_\oto$,
and \Info{oto\_prob\_\allowbreak p($j$)\%prob(0)} is the probability that $M\pj$ has no
counterpart.
\subsection{Simulations}
Mock catalogs are created with the subroutine \Info{simul\_catalogs}
of \Fichier{mod\_simul\_catalogs.f90}.
Here are some details about it.
\subsubsection{Generation of twin mock catalogs}
\label{twin_catalogs}
\Paragraph*{Shape of the surface}
A spherical cap $\mathcal{C}$ covering a solid angle \Info{area\_S}%
\footnote{This parameter is actually optional in \Info{simul\_catalogs}.
  It is set to $4*\piup$ if absent.}
and centered on the North pole is used in simulations.
If $\Info{area\_S}\notin \IOF]0, 4*\piup]*\text{steradians}$, it is set to $4*\piup$ (with a
warning), \ie\ $\mathcal{C}$ is the whole sky.
\Paragraph*{Associations, positions and positional uncertainty angles.
  Side effects}
The procedure described in \S~\paper6.1 has been slightly modified.
The algorithm we now use is the following:
\begin{enumerate}
\item
  Draw the observed positions of all $K'$-sources on $\mathcal{C}$
  randomly and uniformly;
\item
  Generate their true positions from their observed positions
  and from the \Info{semi\_axis\_{\itshape*}\_p} parameters with the subroutine
  \Info{position} of \Fichier{mod\_simul\_catalogs.f90} (\cf~\S~\ref{position}).
  The observed positional uncertainty angles $\beta\pj$ (\ie\ with respect
  to the direction of the North at the observed position) are drawn randomly
  and uniformly in $\IFO[0, \piup[$;
\item
  For each $i$ from $1$ to $n$,
  \begin{enumerate}
  \item
    determine whether $M_i$ is associated to some
    (yet unknown) counterpart in $K'$.
    Do this with probability \Info{input\_f\_u}
    if \Info{force\_ctp\_frac} (\cf~\S~\ref{constants})
    is false, and for $n\times{}$\Info{input\_f\_u}
    of them otherwise.

    If \Info{one\_to\_one} is true, consider $M_i$ as unassociated
    if all the objects in $K'$ are already associated.
    Jump then to point~\ref{unassociated};
  \item
    randomly select a counterpart $M\pj$ to $M_i$ among
    all the $K'$-sources if \Info{one\_to\_one} is false, and
    among all the $K'$-sources which are not associated yet
    if \Info{one\_to\_one} is true;
  \item
    generate the observed position of $M_i$ with subroutine \Info{position}
    from its true position
    (\ie\ that of $M\pj$) and from the \Info{semi\_axis\_{\itshape*}\_u}
    parameters.
    The true positional uncertainty angle (\ie\ with respect
    to the direction of the North at the true position) is drawn randomly
    and uniformly in $\IFO[0, \piup[$, and is converted
    to the observed angle%
    \footnote{%
      Except near the poles, the difference between the true and observed angles
      is negligible.%
    }
    in this subroutine (cf.~\S~\ref{position});
  \item
    \label{side_effects}
    \emph{side effects} may occur if $\mathcal{C}$ is a strict subset
    of the whole sky (\ie\ if $\Info{area\_S} < 4*\piup$):
    a $K'$-source observed within the spherical cap $\mathcal{C}$ but close to
    its boundary may indeed have a counterpart observed outside of
    $\mathcal{C}$.

    If the observed position of $M_i$ falls outside of $\mathcal{C}$,
    discard this position and
    consider $M_i$ as unassociated. Jump then to point~\ref{unassociated};
  \end{enumerate}
\item
  \label{unassociated}
  For each unassociated $K$-source, draw its observed position uniformly
  and randomly on $\mathcal{C}$.
\end{enumerate}
The positions of the sources are ordered by increasing right ascension
only after this procedure.
\subsubsection{Effective fraction of \texorpdfstring{$K$}{K}-sources with a counterpart}
The quantity \Info{eff\_f\_u} may be different from \Info{input\_f\_u}
for several reasons:
\begin{itemize}
\item
  because the number of $K$-sources with a counterpart must be less than~$n'$
  if \Info{one\_to\_one} is true;
\item
  because the number of sources with a counterpart is an integer;
\item
  if $S < 4*\piup$, because of side effects (\cf~point~\ref{side_effects}
  of \S~\ref{twin_catalogs});
\item
  if \Info{force\_ctp\_frac} is false, because
  (ignoring above constraints) \Info{input\_f\_u} is the probability
  that a $K$-source has a counterpart, so \Info{eff\_f\_u} is the same
  as \Info{input\_f\_u} only on average, \ie
  \begin{equation}
    \langle n\times\Info{eff\_f\_u} + \Info{n\_unavailable}
    + \Info{n\_side\_effects}\rangle = n\times\Info{input\_f\_u}.
  \end{equation}

  However, if \Info{force\_ctp\_frac} is true,
  \begin{equation}
    \lvert\mathstrut n\times\Info{eff\_f\_u} + \Info{n\_unavailable}
    + \Info{n\_side\_effects} - n\times\Info{input\_f\_u}\rvert < 1.
  \end{equation}
\end{itemize}
\subsubsection{Using \Info{seed}}
The knowledge of \Info{seed} may be useful in case one of the simulations
(let say the $i$-th) fails.
To generate exactly the same random numbers as in the $i$-th simulation
and to run the latter first,
insert the line
\[
\text{\Info{call random\_seed(put = (/$s_1$,\ $\ldots$,\ $s_q$/))}}
\]
in \Fichier{example\_simul.f90},
just before the call to \Info{simul\_catalog}.
The integers $s_1$, \textellipsis, $s_q$
are the elements of the array \Info{seed} returned for the $i$-th simulation;
do not forget the commas between them, nor the leading \Info{(/} and the
trailing \Info{/)}.
This procedure avoids to run all the simulations~$1$ to~$i-1$ anew just to
search where the problem comes from.
\subsection{Code constants}
\label{constants}
The default type used for reals is given by parameter \Info{real\_type},
which is defined in \Fichier{mod\_types.f90}.

All the following constants are defined and commented in
\Fichier{mod\_constants\allowbreak.f90} (most of them control the accuracy
of the results):
\begin{description}[\textmd{:}]
\item[\Info{force\_ctp\_frac}]
  \leavevmode
  \begin{itemize}
  \item
    \Info{.true.} if the number $n\times\Info{eff\_f\_u}$
    of simulated $K$-sources associated to $K'$-ones is forced
    to be equal to $n\times\Info{input\_f\_u}$, up to one unit;
  \item
    \Info{.false.} if the probability that any $K$-source is associated
    is drawn with probability \Info{input\_f\_u}, independently from the
    association status of other $K$-sources.
  \end{itemize}
  (All this, ignoring side effects and, if \Info{one\_to\_one} is true,
  the constraint $n\times\Info{eff\_f\_u} \le n'$.)

  Default value: \Info{.true.};
\item[\Info{n\_max\_sigma}]
  Value of
  \[
  \frac{\text{\Info{R\_ctp}}}{
    \max_{M_i\in K, M\pj\in K'}\sqrt{a_{\smash[t]{i}}^2 + a_{\smash[t]{j}}'^2}},
  \]
  where \Info{R\_ctp} is the radius within which counterparts
  are searched.
  Reasonable values of \Info{n\_max\_\allowbreak sigma} should be in 
  $\IFF[5, 10]$.

  Default value: $5$;
\item[\Info{min\_xi\_ratio}]
  Minimum value of $\xi_{i, j}/\xi_{i, 0}$, where
  $\xi_{i, j}$ is the spatial probability density to observe the source
  $M\pj$ at some position if it is a true counterpart of the
  object $M_i$, and $\xi_{i, 0}$ is the same quantity if they are
  unrelated.
  This ratio should be $\ll 1$:
  if $\xi_{i, j}/\xi_{i, 0} < \text{\Info{min\_xi\_ratio}}$,
  then $\xi_{i, j}$ is set to $0$ and $M\pj$
  is not considered as a possible counterpart of $M_i$;
  if $\text{\Info{min\_xi\_ratio}} \leqslant 0$,
  this never happens.

  Default value: $10^{-10}$;
\item[\Info{epsilon\_f}]
  Accuracy of maximum-likelihood estimates of $f$ and $f'$:
  an estimate $\hat f$ of $f$ is the limit of a sequence $f_{k+1} = g(f_k)$,
  where $g$ is the function defined in \S~\paper~5.3; when
  $\module{f_{k+1}-f_k} < \text{\Info{epsilon\_f}}$, the
  iteration procedure stops and $\hat f$ is set to $f_{k+1}$.

  Default value: $10^{-5}$.
\end{description}
The constants below are used only in calculations under $H_\oto$:
\begin{description}[:]
\item[\Info{R\_ratio}]
  Value of $\text{\Info{R\_ngb}}/\text{\Info{R\_ctp}}$,
  where \Info{R\_ngb} is the radius within which neighbors
  are searched in calculations under $H_\oto$.
  An \Info{R\_ngb} equal to a few times \Info{R\_ctp} seems
  reasonable.

  Default value: $2$;
\item[\Info{n\_max\_ngb}]
  Maximal number of neighbors (including the main source, so $\geqslant 1$)
  considered for calculations under $H_\oto$.

  Default value: $8$;
\item[\Info{epsilon\_oto\_prob}]
  Parameter setting the accuracy of the iterative procedure computing
  the probabilities $\Poto(A_{i, j} \mid C \cap C')$.
  See subroutine \Info{compute\_oto\_prob} in \Fichier{mod\_Aspects.f90}.

  Default value: $10^{-5}$;
\item[\Info{epsilon\_ln\_oto\_Lh}]
  Parameter setting the accuracy of $\ln\Lh_\oto$ when it is computed with
  \Info{compute\_ln\_\allowbreak oto\_Lh}. See subroutine \Info{refine\_grid}
  in \Fichier{mod\_Aspects.f90}.

  Default value: $10^{-5}$;
\item[\Info{d\_f}]
  Shift of $\hat f_\oto$ used in the subroutine \Info{oto\_analysis}
  of \Fichier{mod\_Aspects.f90} to compute the standard deviation
  of $\hat f_\oto$ from a finite difference approximation to
  $\bigl(\partial^2 \ln\Lh_\oto/\partial f^2\bigr)_{f=\hat f_\oto}$.\vadjust{\vskip-2pt}

  Default value: $10^{-3}$;
\item[\Info{f\_max\_shift}]
  Under $H_\oto$, the theoretical maximal value of $f$ is
  $f_{\text{max}} = \min\{1, n'\!/n\}$.
  For numerical reasons, the maximal value of $f$ effectively used
  in the subroutine \Info{compute\_ln\_oto\_Lh} of \Fichier{mod\_Aspects\allowbreak.f90}
  is $f_{\text{max}} - \text{\Info{f\_max\_shift}}$.

  Default value: $10^{-3}$.
\end{description}
\subsection{Modules used in \Fichier{example\_simul}
  and \Fichier{example\_read}}
A few modules are loaded by \Fichier{example\_simul} and
\Fichier{example\_read}
(see the \Info{use mod\_{\itshape*}} statements at the beginning of these
programs%
\footnote{The commented out \Info{, only :\ \textrm{\itshape\textellipsis}} part
  is just to remind all the variables used.}).
All of them are in the corresponding \Fichier{mod\_{\itshape*}.f90}
files.

The file \Fichier{mod\_variables.f90} gives access to data of interest
for the two programs, in particular the records \Info{assoc\_data\_u}
and \Info{assoc\_data\_p}; these records, of type \Info{assoc\_data}
(see \Fichier{mod\_types.f90}), contain all
the informations on sources (their possible counterparts in the other catalog,
their neighbors in the same catalog, etc.)\ which are necessary to compute
probabilities of association.
Other data which may be used only by \Fichier{example\_simul}
are in \Fichier{mod\_simul\_catalogs.f90};
those only for
\Fichier{example\_read} are in \Fichier{mod\_read\_catalogs.f90}.

The files \Fichier{mod\_Aspects.f90}, \Fichier{mod\_simul\_catalogs.f90},
\Fichier{mod\_read\_\allowbreak catalog.f90} and
\Fichier{mod\_output\_\allowbreak prob.f90} also provide some procedures
which we now describe.
\subsubsection{Procedures defined in \Fichier{mod\_Aspects.f90}}
\begin{description}[\textmd{:}]
\item[\Info{general\_preliminaries}]
  preliminaries required for computations under all assumptions.
  Computes the first part of \Info{assoc\_data\_u} and \Info{assoc\_data\_p};
\item[\Info{oto\_preliminaries}]
  additional preliminaries required for computations under $H_\oto$ only.
  Computes the second part of \Info{assoc\_data\_u} and \Info{assoc\_data\_p};
\item[\Info{sto\_analysis}]
  iteration procedure converging to the
  maximum-likelihood estimate $\hat f_\sto$ of $f$
  (\Info{sto\_f\_u}). Computes also $\hat f'_\sto$
  (\Info{sto\_f\_p})
  and the standard deviation of $\hat f_\sto$
  (\Info{std\_dev\_sto\_f\_u});
\item[\Info{ots\_analysis}]
  iteration procedure converging to the
  maximum-likelihood estimate $\hat f'_\ots$ of $f'$
  (\Info{ots\_\allowbreak f\_p}). Computes also $\hat f_\ots$
  (\Info{ots\_f\_u})
  and the standard deviation of $\hat f'_\ots$
  (\Info{std\_dev\_ots\_f\_p});
\item[\Info{oto\_analysis}]
  iteration procedure converging to the
  maximum-likelihood estimate $\hat f_\oto$ of $f$
  (\Info{oto\_f\_u}); Computes also $\hat f'_\oto$
  (\Info{oto\_f\_p})
  and the standard deviation of $\hat f_\oto$
  (\Info{std\_dev\_oto\_f\_u});
\item[\Info{func\_ln\_sto\_Lh}]
  log-likelihood $\ln \Lh_\sto$ for a given $f$;
\item[\Info{func\_ln\_ots\_Lh}]
  log-likelihood $\ln \Lh_\ots$ for a given $f'$;
\item[\Info{compute\_ln\_oto\_Lh}]
  procedure computing $\ln\Lh_\oto$ for a given $f$, a grid of equidistant values
  of $f$, or both;
\item[\Info{func\_ln\_oto\_Lh\_alt}]
  alternative computation of the log-likelihood $\ln \Lh_\oto$
  for a given $f$. Not recommended;
\item[\Info{compute\_sto\_prob}]
  computes $\Psto(A_{i, j} \mid C \cap C')$ for a given $f$ and for all
  $(i, j) \in \IE[1, n]\times\IE[1, n']\setminus\{(0, 0\}$;
\item[\Info{compute\_ots\_prob}]
  computes $\Pots(A_{i, j} \mid C \cap C')$ for a given $f'$ and for all
  $(i, j) \in \IE[1, n]\times\IE[1, n']\setminus\{(0, 0\}$;
\item[\Info{compute\_oto\_prob}]
  computes $\Poto(A_{i, j} \mid C \cap C')$ for a given $f$
  and for all $(i, j) \in \IE[1, n]\times\IE[1, n']\setminus\{(0, 0\}$.
\end{description}
\subsubsection{Procedures defined in \Fichier{mod\_simul\_catalogs.f90}}
\begin{description}[:]
\item[\Info{simul\_catalogs}] creates twin mock catalogs;
\item[\Info{write\_catalogs}] writes twin catalogs to output files;
\item[\Info{seed\_format}] output format for the array of seeds used by
  the generator of random numbers.
\end{description}
\subsubsection{Input/output procedures}
\Paragraph*{Defined in \Fichier{mod\_read\_catalog.f90}}
\begin{description}[:]
\item[\Info{read\_catalog}] reads a single catalog.
\end{description}
\Paragraph*{Defined in \Fichier{mod\_output\_prob.f90}}
\begin{description}[:]
\item[\Info{output\_sto\_prob}, \Info{output\_ots\_prob},
  \Info{output\_oto\_prob}]
  writes in files the probabilities of association to every source in $K$ and
  $K'$
  of all its possible counterparts;
\item[\Info{write\_prob}]
  writes in a file the probabilities of association to a single source
  of all its possible counterparts.
  It is called indirectly in \Fichier{example\_read} through the
  \Info{output\_{\itshape*}t{\itshape*}\_prob} procedures described above,
  but may also be called directly.
\end{description}
%%%%%%%%%%%%%%%%%%%%%%%%%%%%%%%%%%%%%%%%%%%%%%%%%%%%%%%%%%%%%%%%%%%%%% 
\newpage
\begin{appendices}
  \centerline{\bfseries\LARGE Complements}
  \addtocontents{toc}{\bigskip\bigskip\centerline{\bfseries\large Complements}\medskip}
  \vspace*{1cm}
  %%%%%%%%%%%%%%%%%%%%%%%%%%%%%%%%%%%%%%%%%%%%%%%%%%%%%%%%%%%%%%%%%%%%%% 
  \section{Efficient search for possible counterparts}
  The search procedure for possible counterparts
  described in \S~\paper5.2 may be accelerated in the following way.
  For most $M_i$, one does not need to \emph{test} for each $M\pk$
  whether $\psi_{i, k} \leqslant R'$.
  Let us write $E_i$ the domain of right ascensions $\alpha'$ out of which
  no point $M'$ of declination $\delta'$ and closer to $M_i$ than
  distance~$R'$ may be found.
  The angular distance $\psi(M_i, M')$ between $M'$ and $M_i$ is given
  (cf.~Eq.~\paper A.15) by
  \begin{equation}
    \cos\psi(M_i, M')
    = \cos(\alpha'-\alpha_i)*\cos\delta_i*\cos\delta' + \sin\delta_i*\sin\delta'.
  \end{equation}

  If $\delta_i \not\in [-\piup/2+R', \piup/2-R']$, then $E_i = [0, 2*\piup]$.
  Else,
  the minimum of $\cos(\alpha'-\alpha_i)$
  under the constraint $\cos\psi(M_i, M') \geqslant \cos R'$ is reached
  when $\sin\delta' = \sin\delta_i/{\cos R'}$ and
  \begin{equation}
    \cos(\alpha'-\alpha_i)
    = \cos\Delta_i,
    \quad \text{where }
    \Delta_i
    =
    \arccos
    \frac{\!\sqrt{\cos^2 R' - \sin\mathclose{}^2\,\delta_i}}{\cos\delta_i}.
  \end{equation}
  The domain of possible right ascensions is then given by
  \begin{equation}
    E_i = \left\{
      \begin{aligned}
        &[0, \alpha_i + \Delta_i - 2*\piup] \cup
        [\alpha_i-\Delta_i, 2*\piup]
        && \text{if } \alpha_i + \Delta_i > 2*\piup,
        \\
        &[0, \alpha_i + \Delta_i] \cup
        [\alpha_i-\Delta_i + 2*\piup, 2*\piup]
        && \text{if } \alpha_i - \Delta_i < 0,
        \\
        &[\alpha_i - \Delta_i, \alpha_i+\Delta_i]
        && \text{otherwise}.
      \end{aligned}
    \right.
  \end{equation}

  For a catalog $K'$ ordered by increasing right ascension
  (if not, this is the first thing to do),
  one may easily find the subset
  of indices $k$ for which $\alpha\pk \in E_i$.
  For instance,
  if $E_i = [\alpha_i - \Delta_i, \alpha_i+\Delta_i]$,
  one just has to find by dichotomy the indices $k^-$ and $k^+$ such that
  $\alpha'_{\smash[t]{k^--1}} < \alpha_i - \Delta_i
  \leqslant \alpha'_{\smash[t]{k^-}}$
  and $\alpha'_{\smash[t]{k^+}} \leqslant \alpha_i + \Delta_i
  < \alpha'_{\smash[t]{k^++1}}$;
  sums like
  $\sum_{k=1{;}\, \psi_{i, k}\leqslant R'}^{n'}$ may then be replaced by
  $\sum_{k=k^-{;}\, \psi_{i, k}\leqslant R'}^{k^+}$.

  Whatever the domain $E_i$,
  these sums may be further restricted to sources with a declination
  $\delta'_{\smash[t]{k}} \in [\delta_i-R',\allowbreak \delta_i+R'] \cap [-\piup/2, \piup/2]$.

  This procedure is implemented in subroutine \Info{alpha\_bounds}
  of \Fichier{mod\_Aspects.f90}.
  % 
  %%%%%%%%%%%%%%%%%%%%%%%%%%%%%%%%%%%%%%%%%%%%%%%%%%%%%%%%%%%%%%%%%%%%%% 
  \section{Simulation of positions}
  \label{position}
  We provide here some explanations on how subroutine \Info{position}
  of file \Fichier{mod\_simul\_catalogs.f90} simulates
  the observed position of a source
  from its true position (or the converse) and its positional uncertainty
  parameters.

  Let $M_0(\alpha_0, \delta_0)$ be the true position of a source,
  $a$ ($\ll \piup$~radians) and $b$
  be the semi-major and semi-minor axes of the positional uncertainty ellipse,
  and $\vu_a$ and $\vu_b$ be unit vectors along the major and minor axes
  ($\vu_a$ and $\vu_b$ are oriented so
  that $(\vu_a, \vu_b, \vu_{r_0})$ is a direct orthonormal basis).

  For any vectors $\vec v\,$ and $\vec w$ perpendicular to $\vu_{r_0}$, we
  denote by $\pangle{\vec v}{\vec w}$ the angle from $\vec v\,$ to $\vec w$
  oriented counter\-clock\-wise around $\vu_{r_0}$, and by
  $\mangle{\vec v}{\vec w}$ the same angle oriented clockwise
  (\ie,
  $\mangle{\vec v}{\vec w} = -\pangle{\vec v}{\vec w} = \pangle{\vec w}{\vec v}$).

  We want to simulate an observed position, $M(\alpha, \delta)$.
  First, define $\vec n \egdef  \vu_{r_0}\times\vu_r/\|\vu_{r_0}\times\vu_r\|$,
  $\vec t \egdef  \vec n\times\vec u_{r_0}$ (the basis $(\vec t, \vec n, \vec u_{r_0})$
  is therefore direct and orthonormal) and $\psi \egdef  \angle(\vu_{r_0}, \vu_r)$.

  Let $\beta_0 \egdef  \mangle{\vu_{\delta_0}}{\vu_a}$ and
  $\beta \egdef  \mangle{\vu_\delta}{\vu_a}$ be the true and observed
  positional uncertainty angles, and define
  $\gamma_0 \egdef  \mangle{\vec n}{\vu_{\delta_0}}$,
  $\gamma \egdef  \mangle{\vec n}{\vu_\delta}$
  and $\epsilon \egdef  \pangle{\vu_a}{\vec t\,}$
  ($\gamma$ and $\beta$ are in the plane perpendicular
  to $\vu_r$, not to $\vu_{r_0}$, but one may consider here that
  $\vu_r \approx \vu_{r_0}$; more
  rigorously, one may project the sphere on the tangent cylinder containing
  $M_0$ and $M$ and unroll this cylinder on a plane).

  The following relations hold:
  \begin{equation}
    \gamma_0 + \beta_0 = \gamma + \beta = \mangle{\vec n}{\vu_a},
  \end{equation}
  so
  \vspace{-\abovedisplayskip}\vspace{\abovedisplayshortskip}%
  \begin{align}
    \gamma_0 + \beta_0 - \epsilon & = \gamma + \beta - \epsilon 
    \notag\\
    &= \mangle{\vec n}{\vu_a} - \pangle{\vu_a}{\vec t\,}
    = \mangle{\vec n}{\vu_a} + \mangle{\vu_a}{\vec t\,} 
    = \pangle{\vec t\,}{\vec n}
    \notag\\
    &= \piup/2 \pmod{2*\piup};
    \label{beta_gamma}\\
    \cos\psi &= \cos\delta_0*\cos\delta*\cos(\alpha-\alpha_0)
    + \sin\delta_0*\sin\delta; 
    \label{cospsi}\\
    \cos\gamma_0 &= \frac{\cos\delta*\sin(\alpha-\alpha_0)}{\sin\psi};
    \label{cosgammaz}\\
    \sin\gamma_0 &= \frac{\cos\delta_0*\sin\delta -
      \sin\delta_0*\cos\delta*\cos(\alpha-\alpha_0)}{\sin\psi}.
    \label{singammaz}
  \end{align}
  (Cf.~\S~\paper A for Eqs.~\eqref{cospsi} to~\eqref{singammaz}.)

  One has
  \begin{equation}
    \overrightarrow{M_0M} \approx \psi*\vec t
    = \psi*(\cos\epsilon*\vu_a + \sin\epsilon*\vu_b)
    = a*r_1*\vu_a+b*r_2*\vu_b,
  \end{equation}
  where $r_1$ and $r_2$ are independent random numbers drawn from a normal
  distribution with mean~$0$ and standard deviation~$1$.
  Therefore, $\psi \approx ([a*r_1]^2+[b*r_2]^2)^{1/2}$ and,
  using the Fortran function \Info{atan2} to obtain a unique result in
  $\IOF]-\piup, \piup]$, one obtains
  $\epsilon \approx \Info{atan2}(b*r_2, a*r_1)$.

  One may either draw $\beta_0$ randomly and derive $\alpha$, $\delta$ and
  $\beta$ (as in the code; \S~\ref{beta}), or draw $\beta$ randomly and derive
  $\alpha$ and $\delta$ directly (\S~\ref{beta0}; not implemented).
  Note that in the second
  case, two positions may be valid solutions of the set of equations.
  \subsection{\texorpdfstring{Determination of $\alpha$, $\delta$ and $\beta$
    from $\beta_0$ (and from $\alpha_0$, $\delta_0$, $\psi$
    and $\epsilon$)}{Determination of \textbackslash alpha, \textbackslash delta and \textbackslash beta
    from \textbackslash beta\_0 (and from \textbackslash alpha\_0, \textbackslash delta\_0, \textbackslash psi
    and \textbackslash epsilon)}}
  \label{beta}
  Angle $\gamma_0$ is first derived from Eq.~\eqref{beta_gamma}, $\beta_0$ and
  $\epsilon$.

  Combining Eq.~\eqref{singammaz} and Eq.~\eqref{cospsi}, one gets first
  \begin{equation}
    \sin\gamma_0*\sin\psi = \cos\delta_0*\sin\delta -
    \sin\delta_0*\frac{\cos\psi - \sin\delta_0*\sin\delta}{\cos\delta_0},
  \end{equation}
  then
  \begin{equation}
    \delta = \arcsin(\sin\gamma_0*\sin\psi*\cos\delta_0 +
    \sin\delta_0*\cos\psi).
  \end{equation}

  From Eq.~\eqref{singammaz} and Eq.~\eqref{cosgammaz}, one obtains
  \begin{equation}
    \sin(\alpha-\alpha_0) = \frac{\cos\gamma_0*\sin\psi}{\cos\delta}
  \end{equation}
  and
  \begin{equation}
    \cos(\alpha-\alpha_0) = \frac{\cos\delta_0*\sin\delta -
      \sin\gamma_0*\sin\psi}{\sin\delta_0*\cos\delta}.
  \end{equation}
  Thus,
  \begin{equation}
    \alpha = \alpha_0 +
    \Info{atan2}(\sin[\alpha-\alpha_0], \cos[\alpha-\alpha_0]) \pmod{2*\piup}.
  \end{equation}

  Using \Info{atan2}, one derives $\gamma$ from
  \begin{equation}
    \label{cosgamma}
    \cos\gamma = \frac{\cos\delta_0*\sin(\alpha-\alpha_0)}{\sin\psi}
  \end{equation}
  and
  \begin{equation}
    \label{singamma}
    \sin\gamma = \frac{\cos\delta_0*\sin\delta*\cos(\alpha-\alpha_0) -
      \sin\delta_0*\cos\delta}{\sin\psi},
  \end{equation}
  and then $\beta$ from Eq.~\eqref{beta_gamma} and $\epsilon$.
  \subsection{\texorpdfstring{Determination of $\alpha$, $\delta$ from $\beta$ 
    (and from $\alpha_0$, $\delta_0$, $\psi$ and $\epsilon$)}{Determination of \textbackslash alpha, \textbackslash delta from \textbackslash beta 
    (and from \textbackslash alpha\_0, \textbackslash delta\_0, \textbackslash psi and \textbackslash epsilon)}}
  \label{beta0}
  Angle $\gamma$ is first derived from Eq.~\eqref{beta_gamma}, $\beta$ and
  $\epsilon$.

  From Eq.~\eqref{cosgamma}, one obtains
  \begin{equation}
    \alpha = \alpha^+ =
    \alpha_0 + \arcsin\frac{\cos\gamma * \sin\psi}{\cos\delta_0} \pmod{2*\piup}
  \end{equation}
  or
  \begin{equation}
    \alpha = \alpha^- = \alpha_0 + \piup -
    \arcsin\frac{\cos\gamma * \sin\psi}{\cos\delta_0} \pmod{2*\piup}.
  \end{equation}

  The combination
  $\sin\delta_0 \times \text{Eq.}~\eqref{cospsi} +
  \cos\delta_0 * \cos(\alpha^{\pm}-\alpha_0) *\sin\psi \times
  \text{Eq.}~\eqref{singamma}$
  leads to
  \begin{equation}
    \sin\delta^{\pm} = \frac{\sin\delta_0 * \cos\psi +
      \cos\delta_0 * \cos(\alpha^{\pm}-\alpha_0) * \sin\gamma * \sin\psi}{
      1 - \cos^2\gamma * \sin^2\psi},
  \end{equation}
  and $\cos\delta_0 * \cos(\alpha^{\pm}-\alpha_0) \times
  \text{Eq.}~\eqref{cospsi} - \sin\delta_0 * \sin\psi \times
  \text{Eq.}~\eqref{singamma}$ to
  \begin{equation}
    \cos\delta^{\pm} =
    \frac{\cos\psi * \cos\delta_0 * \cos(\alpha^{\pm}-\alpha_0) -
      \sin\gamma * \sin\psi * \sin\delta_0}{1 - \cos^2\gamma * \sin^2\psi}.
  \end{equation}
  Declinations $\delta^+$ and $\delta^-$ are then derived using
  \Info{atan2}.
  Values of $\delta^{\pm}$ out of $\IFF[-\piup/2, \piup/2]$ are excluded, but both
  may fall in this interval: one must arbitrarily select one of the
  positions $(\alpha^+, \delta^+)$ and $(\alpha^-, \delta^-)$.
  % 
  %%%%%%%%%%%%%%%%%%%%%%%%%%%%%%%%%%%%%%%%%%%%%%%%%%%%%%%%%%%%%%%%%%%%%% 
  \section{Computation of \texorpdfstring{$\Lh_\oto$}{L\_\{o:o\}}}
  \subsection{Through integration of \texorpdfstring{$\partial\ln\Lh_\oto/\partial f$}{\textbackslash partial ln L\_\{o:o\}/\textbackslash partial f}}
  The most efficient way to compute $\Lh_\oto$ is to integrate
  Eq.~(\paper80) with respect to $f$.
  This is done with the adaptative quadrature subroutine
  \Info{compute\_ln\_oto\_Lh} of \Fichier{mod\_Aspects.f90}.
  This procedure may compute $\ln\Lh_\oto$ for a single $f$,
  a grid of equidistant values of $f$ starting at $0$, or both.

  The integration starts at $f = 0$.
  Subroutine \Info{compute\_ln\_oto\_Lh} first computes
  $d(f) \egdef \partial\ln\Lh_\oto/\partial f$
  (function \Info{drv\_ln\_oto\_Lh}) for a regular grid of values
  of $f$.
  Because $d$ may vary a lot on a bin,
  especially near $f = 0$, the grid is then refined
  (procedure \Info{refine\_grid}):
  each bin $\IFF[f_{\text{i}}, f_{\text{s}}]$ is recursively subdivided
  until
  \begin{equation}
    \Valabs{d(f_{\text{i}}) + d(f_{\text{s}})
      - 2*d(f_{\text{m}})} < \Info{epsilon\_ln\_oto\_Lh} \times
    \left(\Valabs{d[f_{\text{i}}]} + \Valabs{d[f_{\text{s}}]}
      + 2*\Valabs{d[f_{\text{m}}]}\right),
  \end{equation}
  where $f_{\text{m}} = (f_{\text{i}} + f_{\text{s}})/2$
  and \Info{epsilon\_ln\_oto\_Lh} ($> 0$ and $\ll 1$) is a parameter defined in
  \Fichier{mod\_constants.f90} (\cf~\S~\ref{constants}).
  This ensures that the curvature of $d$ in $\IFF[f_{\text{i}}, f_{\text{s}}]$
  is negligible and that the integration of $d$ with the trapezoidal rule
  (\ie, interpolating $d$ on the bin with a linear function) would provide an
  accurate value of
  $\ln\Lh_\oto(f_{\text{s}}) - \ln\Lh_\oto(f_{\text{i}})$ (typically, at most
  of the order of \Info{epsilon\_ln\_oto\_Lh} in relative value).

  Actually, instead of linear functions, we use cubic functions.
  The latter are interpolated on the values of $d$
  in the bin of interest and the two adjacent ones, following the procedure
  described in \citet{Steffen} (function \Info{Steffen\_integral}).
  The final accuracy of $\ln\Lh_\oto$ should therefore be much better than
  with the trapezoidal rule.
  %%%%%%%%%%%%%%%%%%%%%%%%%%%%%%%%%%%%%%%%%%%%%%%%%%%%%%%%%%%%%%%%%%%%%% 
  \subsection{Alternative method}
  \label{other_Lh_oto}
  The one-to-one likelihood may be computed in another way than
  from Eqs.~(\paper80) and~(\paper82)
  (cf.~\S~\paper4.2).
  To do this, we first define
  \begin{equation}
    \Lambda_i \coloneqq \Poto\Bigl(\bigcap_{k=i}^n c_k \Bigm| C' \cap
    \bigcap_{k=1}^{i-1}A_{k, 0}\Bigr).
  \end{equation}
  One has
  \begin{equation}
    \label{Lh_oto_0}
    \Poto(C \mid C')
    = \Lambda_1
  \end{equation}
  and, for any $i \in \IE[1, n]$, $\Lambda_i$ may be computed as a function of
  $\Lambda_{i+1}$ in the following way.

  For $\omega_1 = A_{i, 0}$,
  $\omega_2 = \bigcap_{\smash[t]{k=i}}^n c_k$
  and $\omega_3 = C' \cap \bigcap_{\smash[t]{k=1}}^{i-1} A_{k, 0}$,
  Eq.~(\paper3) gives
  \begin{equation}
    \label{Lh_oto_1}
    \Lambda_i
    =
    \frac{
      \Poto(A_{i, 0} \cap \bigcap_{k=i}^n c_k
      \mid C' \cap \bigcap_{k=1}^{i-1} A_{k, 0})
    }{
      \Poto(A_{i, 0} \mid [\bigcap_{k=i}^n c_k]
      \cap C' \cap \bigcap_{k=1}^{i-1} A_{k, 0})
    }.
  \end{equation}

  Let us compute the numerator of Eq.~\eqref{Lh_oto_1}.
  With $\omega_1 = c_i$,
  $\omega_2 = \bigcap_{\smash[t]{k=i+1}}^n c_k \cap A_{i, 0}$
  and $\omega_3 = C' \cap \bigcap_{\smash[t]{k=1}}^{i-1} A_{k, 0}$
  in Eq.~(\paper3), one obtains
  \begin{equation}
    \label{Lh_oto_2}
    \Poto\Bigl(A_{i, 0} \cap \bigcap_{k=i}^n c_k
    \Bigm| C' \cap \bigcap_{k=1}^{i-1} A_{k, 0}\Bigr)
    =
    \Poto\Bigl(c_i \Bigm| \Bigl[\bigcap_{k=i+1}^n c_k\Bigr] \cap C'
    \cap \bigcap_{k=1}^i A_{k, 0}\Bigr)
    *
    \Poto\Bigl(\bigcap_{k=i+1}^n c_k \cap A_{i, 0}
    \Bigm| C' \cap \bigcap_{k=1}^{i-1} A_{k, 0}\Bigr).
  \end{equation}
  The first factor in the right-hand side of Eq.~\eqref{Lh_oto_2} is
  simply
  \begin{equation}
    \label{Lh_oto_2_first}
    \Poto\Bigl(c_i
    \Bigm| \Bigl[\bigcap_{k=i+1}^n c_k\Bigr] \cap C'
    \cap \bigcap_{k=1}^i A_{k, 0}\Bigr)
    = \Poto(c_i \mid A_{i, 0})
    = \xi_{i, 0}*\df^2\vec r_i.
  \end{equation}
  The second factor is
  \begin{equation}
    \label{Lh_oto_2_second}
    \Poto\Bigl(\bigcap_{k=i+1}^n c_k \cap A_{i, 0}
    \Bigm| C' \cap \bigcap_{k=1}^{i-1} A_{k, 0}\Bigr)
    =
    \Poto\Bigl(\bigcap_{k=i+1}^n c_k
    \Bigm| C' \cap \bigcap_{k=1}^i A_{k, 0}\Bigr)
    *
    \Poto\Bigl(A_{i, 0} \Bigm| C' \cap \bigcap_{k=1}^{i-1} A_{k, 0}\Bigr),
  \end{equation}
  where we have used once again Eq.~(\paper3),
  with $\omega_1 = \bigcap_{\smash[t]{k=i+1}}^n c_k$,
  $\omega_2 = A_{i, 0}$
  and $\omega_3 = C' \cap \bigcap_{\smash[t]{k=1}}^{i-1} A_{k, 0}$.

  The right-most term in Eq.~\eqref{Lh_oto_2_second} is
  \begin{equation}
    \label{Lh_oto_2_second_last}
    \Poto\Bigl(A_{i, 0}
    \Bigm| C' \cap \bigcap_{k=1}^{i-1} A_{k, 0}\Bigr)
    = \Poto(A_{i, 0}) = 1-f,
  \end{equation}
  so, combining Eqs.~\eqref{Lh_oto_1}, \eqref{Lh_oto_2},
  \eqref{Lh_oto_2_first}, \eqref{Lh_oto_2_second}, \eqref{Lh_oto_2_second_last}
  and~(\paper19),
  one obtains
  \begin{equation}
    \label{Lh_oto_3}
    \Lambda_i
    =
    \frac{
      \zeta_{i, 0}*\df^2\vec r_i
    }{
      \Poto(A_{i, 0} \mid [\bigcap_{k=i}^n c_k]
      \cap C' \cap \bigcap_{k=1}^{i-1} A_{k, 0})
    }
    *
    \Lambda_{i+1}.
  \end{equation}

  From Eq.~\eqref{Lh_oto_3},
  we get by iteration that
  \begin{equation}
    \Lambda_1
    =
    \Biggl(\prod_{i=1}^n \frac{
      \zeta_{i, 0}*\df^2\vec r_i
    }{
      \Poto[A_{i, 0} \mid (\bigcap_{k=i}^n c_k)
      \cap C' \cap \bigcap_{k=1}^{i-1} A_{k, 0}]
    }\Biggr)*\Lambda_{n+1}
  \end{equation}
  and, since $\Poto(C \mid C') = \Lambda_1$ and $\Lambda_{n+1} = 1$, that
  \begin{equation}
    \label{P_oto(C|C')_iter}
    \Poto(C \mid C')
    =
    \prod_{i=1}^n \frac{
      \zeta_{i, 0}*\df^2\vec r_i
    }{
      \Poto(A_{i, 0} \mid [\bigcap_{k=i}^n c_k]
      \cap C' \cap \bigcap_{k=1}^{i-1} A_{k, 0})
    }.
  \end{equation}

  Because the event $A_{i, 0}$ does not depend on the coordinates
  $\bigcap_{\smash[t]{k=1}}^{i-1} c_k$ of sources without counterpart,
  \begin{equation}
    \label{P_oto(Ai0|C,C',Ak0,k<i)}
    \Poto\Bigl(A_{i, 0} \Bigm| \Bigl[\bigcap_{k=i}^n c_k\Bigr]
    \cap C' \cap \bigcap_{k=1}^{i-1} A_{k, 0}\Bigr)
    =
    \Poto\Bigl(A_{i, 0} \Bigm| C \cap C'
    \cap \bigcap_{k=1}^{i-1} A_{k, 0}\Bigr).
  \end{equation}
  Finally, one obtains from Eqs.~\eqref{P_oto(C|C')_iter},
  \eqref{P_oto(Ai0|C,C',Ak0,k<i)}, (\paper31) and~(\paper13)
  that
  \begin{equation}
    \label{Lh_oto_iter}
    \Lhoto =  \Biggl(\prod_{i=1}^n \frac{
      \zeta_{i, 0}
    }{
      \Poto[A_{i, 0} \mid C \cap C' \cap
      \bigcap_{k=1}^{i-1} A_{k, 0}]
    }\Biggr)*\prod_{j=1}^{n'} \xi_{0, j}.
  \end{equation}
  This formula is actually also valid
  for $\Lhsto$ (with $\Poto$ replaced by $\Psto$),
  but is much less convenient than Eq.~(\paper32).

  The terms $\Poto(A_{i, 0}
  \mid C \cap C' \cap \bigcap_{\smash[t]{k=1}}^{i-1} A_{k, 0})$\vadjust{\vskip-2pt}
  in Eq.~\eqref{Lh_oto_iter} can be computed in the
  same way as the probabilities
  $\Poto(A_{i, j} \mid C \cap C')$
  (cf.~\S~\paper5.4):
  to take into account the constraint
  $\bigcap_{\smash[t]{k=1}}^{i-1} A_{k, 0}$,
  one may for instance restrict
  to $j_k = 0$ the sums on $j_k$ in Eq.~(\paper96), and this
  for all indices $k$ such that $\phi(k) < i$.

  This method is used in function \Info{func\_ln\_oto\_Lh\_alt}
  of \Fichier{mod\_Aspects.f90}.
  The current implementation is however very inefficient and
  the subroutine \Info{compute\_ln\_oto\_Lh} should be preferred.
  %%%%%%%%%%%%%%%%%%%%%%%%%%%%%%%%%%%%%%%%%%%%%%%%%%%%%%%%%%%%%%%%%%%%%% 
  \section{One-to-one computations for \texorpdfstring{$n > n'$}{n > n'}}
  \label{larger}
  Although one always can define $K$ and $K'$
  under assumption~$H_\oto$
  in such a way that $n \leqslant n'$, it is interesting to treat the case
  where $n > n'$ to check the consistency of numerical calculations
  (cf.~\S~\paper5.5).
  \subsection{Probability of association}
  If $n > n'$, then $n$ and $f$ are replaced by $n'$ and $f'$ in
  Eq.~(\paper61).
  The denominator
  (Eq.~(\paper65)) of Eq.~(\paper4)
  becomes
  \begin{equation}
    \Poto(C \mid C') =
    \Xi*\sum_{\substack{j_1=0\\ j_1\not\in X_0}}^{n'}
    \sum_{\substack{j_2=0\\ j_2\not\in X_1}}^{n'}
    \cdots \sum_{\substack{j_n=0\\ j_n\not\in X_{n-1}}}^{n'}
    {\frac{(n-q)!}{n!}
      * f'^q * (1-f')^{n'-q}*\prod_{k=1}^n \xi_{k, j_k}}
    = \Xi*(1-f')^{n'-n}*
    \sum_{\substack{j_1=0\\ j_1\not\in X_0}}^{n'}
    \sum_{\substack{j_2=0\\ j_2\not\in X_1}}^{n'}\cdots
    \sum_{\substack{j_n=0\\ j_n\not\in X_{n-1}}}^{n'}
    \prod_{k=1}^n \eta'_{\smash[t]{k, j_k}}\,,
  \end{equation}
  where
  \begin{equation}
    \eta'_{\smash[t]{k, 0}} \coloneqq (1-f')*\xi_{k, 0}
    \qquad\text{and}\qquad
    \eta'_{\smash[t]{k, j_k}} \coloneqq \frac{f'*\xi_{k, j_k}}{n-\card X_{k-1}}
    \quad\text{if } j_k \neq 0.
  \end{equation}

  The numerator (Eq.~(\paper69)) is similarly modified:
  \begin{equation}
    \Poto(A_{i, j} \cap C \mid C')
    =
    \Xi*(1-f')^{n'-n}*\zeta'_{\smash[t]{i, j}}
    \sum_{\substack{j_1=0\\ j_1\not\in X^\star_0}}^{n'} \cdots
    \sum_{\substack{j_{i-1}=0\\ j_{i-1}\not\in X^\star_{i-2}}}^{n'}
    \sum_{\substack{j_{i+1}=0\\ j_{i+1}\not\in X^\star_{i}}}^{n'} \cdots
    \sum_{\substack{j_n=0\\ j_n\not\in X^\star_{n-1}}}^{n'}
    \prod_{\substack{k=1\\ k\neq i}}^n \eta'^\star_{k, j_k}\,,
  \end{equation}
  where
  \begin{gather}
    \zeta'_{\smash[t]{i, 0}} \coloneqq (1-f')*\xi_{i, 0}
    \qquad\text{and}\qquad
    \zeta'_{\smash[t]{i, j}} \coloneqq \frac{f'*\xi_{i, j}}n
    = \frac{f*\xi_{i, j}}{n'}
    \quad\text{if } j \neq 0,
    \\
    \eta'^\star_{k, 0} \coloneqq (1-f')*\xi_{k, 0}
    \qquad\text{and}\qquad
    \eta'^\star_{k, j_k} \coloneqq \frac{f'*\xi_{k, j_k}}{n-\card X^\star_{k-1}}
    \quad\text{if } j_k \neq 0.
  \end{gather}
  \subsection{Expression of \texorpdfstring{$\partial\ln\Lhoto/\partial x_p$}{\textbackslash partial ln L\_\{o:o\}/\textbackslash partial x\_p}}
  Eq.~(\paper76) becomes
  \begin{equation}
    \begin{split}
      \frac{\partial \Poto(C \mid C')}{\partial x_p}
      & =
      \Xi*\frac{\partial\bigl([1-f']^{n'-n}\bigr)}{\partial x_p}*
      \sum_{\substack{j_1=0\\ j_1\not\in X_0}}^{n'}
      \sum_{\substack{j_2=0\\ j_2\not\in X_1}}^{n'}\cdots
      \sum_{\substack{j_n=0\\ j_n\not\in X_{n-1}}}^{n'}
      \prod_{k=1}^n \eta'_{\smash[t]{k, j_k}} \\
      & \phantom{={}}+
      \Xi * (1-f')^{n'-n} *
      \sum_{\substack{j_1=0\\j_1\notin X_0}}^{n'}\sum_{\substack{j_2=0\\j_2\notin X_1}}^{n'}
      \cdots\sum_{\substack{j_n=0\\j_n\notin X_{n-1}}}^{n'}\sum_{i=1}^n{
        \frac{\partial\ln\eta'_{\smash[t]{i, j_i}}}{\partial x_p}*
        \prod_{k=1}^n\eta'_{\smash[t]{k, j_k}}}\,,
    \end{split}
  \end{equation}
  so
  \begin{equation}
    \frac{\partial\ln\Lhoto}{\partial x_p}
    =
    (n'-n)*\frac{\partial\ln(1-f')}{\partial x_p}
    +
    \sum_{i=1}^n\sum_{j=0}^{n'}{\frac{\partial\ln\zeta'_{\smash[t]{i, j}}}{
        \partial x_p}* \Poto(A_{i, j} \mid C \cap C')}.
  \end{equation}
  For $x_p \neq f$, one recovers Eq.~(\paper79) with $\zeta'$ instead of $\zeta$.

  For $x_p = f$ ($= n'*f'/n)$, Eq.~(\paper80) becomes
  \begin{equation}
    \frac{\partial\ln\Lhoto}{\partial f}
    = \frac{
      n* (1-f) - \sum_{i=1}^n \Poto(A_{i, 0} \mid C \cap C')
    }{
      f* (1-f')
    }.
  \end{equation}
  The numerator being the same, Eq.~(\paper81) still holds.
  \subsection{Alternative expression of the likelihood}
  The alternative expression of the likelihood given by Eq.~\eqref{Lh_oto_iter}
  becomes
  \begin{equation}
    \Lhoto = (1-f')^{n'-n}*
    \Biggl(\prod_{i=1}^n \frac{
      \zeta'_{\smash[t]{i, 0}}
    }{
      \Poto[A_{i, 0} \mid C \cap C'
      \cap \bigcap_{k=1}^{i-1} A_{k, 0}]} \Biggr)*
    \prod_{j=1}^{n'} \xi_{0, j}\,,
  \end{equation}
  where
  $\Poto(A_{i, 0} \mid C \cap C'
  \cap \bigcap_{k=1}^{i-1} A_{k, 0})$
  is computed as detailed in \S~\paper5.4.3 from Eq.~(\paper96),
  with $\zeta$ replaced by $\zeta'$ and with $f'$ and $n$ instead
  of $f$ and $n'$
  in the expressions of $\widetilde\eta$ and $\widetilde\eta^{\,\star}$.
  %%%%%%%%%%%%%%%%%%%%%%%%%%%%%%%%%%%%%%%%%%%%%%%%%%%%%%%%%%%%%%%%%%%%%% 
  \section{One-to-several results}
  \label{app_ots}
  The one-to-several assumption
  is fully symmetrical to $H_\sto$, and all results derived under
  $H_\ots$ may be obtained from \S~\paper3 by
  swapping $K$ and $K'$.
  We give them here for convenience.

  The probabilities of association are
  \begin{equation}
    \label{P_ots}
    \Pots(A_{i, j} \mid C \cap C') =
    \left\{
      \begin{aligned}
        \frac{f'*\xi_{i, j}}{
          (1-f')*n*\xi_{0, j} + f'*\sum_{k=1}^n\xi_{k, j}}
        & \quad \text{if } i \neq 0,
        \\
        \frac{(1-f')*n*\xi_{0, i}}{
          (1-f')*n*\xi_{0, j} + f'*\sum_{k=1}^n\xi_{k, j}}
        & \quad \text{if } i = 0
      \end{aligned}
    \right.
  \end{equation}
  for $j\neq 0$ (cf.~Eq.~(\paper24)),
  and
  \begin{equation}
    \Pots(A_{i, 0} \mid C \cap C') =
    \prod_{j=1}^{n'}{(1 - \Pots[A_{i, j} \mid C \cap C'])}
    \qquad\text{(cf.~Eq.~(\paper26))}.
  \end{equation}

  The overall likelihood is
  \begin{equation}
    \label{Lh_ots}
    \Lhots = \Bigl(\prod_{i=1}^n \xi_{i, 0}\Bigr)*
    \prod_{j=1}^{n'}{\Biggl([1-f']*\xi_{0, j} +
      \frac{f'}n*\sum_{k=1}^n\xi_{k, j}\Biggr)}
    \qquad\text{(cf.~Eq.~(\paper32))}.
  \end{equation}

  A maximum likelihood estimator of $f'$ is
  \begin{equation}
    \label{f'_ots}
    \expandafter\hat f'_\ots = 1 - \frac{1}{n'}*\sum_{j=1}^{n'}
    \expandafter\hat\Pots(A_{0, j} \mid C \cap C')
    \qquad\text{(cf.~Eq.~(\paper39))},
  \end{equation}
  where $\expandafter\hat\Pots$ is the value of $\Pots$ at
  $f' = \expandafter\hat f'_\ots$.

  An estimator of $f$ is
  \begin{equation}
    \label{f_ots}
    \hat f_\ots = 1 - \frac{1}n*
    \sum_{i=1}^n\expandafter\hat\Pots(A_{i, 0}
    \mid C \cap C')
    \qquad\text{(cf.~Eq.~(\paper42))}.
  \end{equation}
  %%%%%%%%%%%%%%%%%%%%%%%%%%%%%%%%%%%%%%%%%%%%%%%%%%%%%%%%%%%%%%%%%%%%%% 
  \section{Tedious several-to-one calculations}
  \subsection{Expression of \texorpdfstring{$\partial^2\ln\Lh_\sto/\partial f^2$}{\textbackslash partial\textasciicircum2 ln L\_\{s:o\}/\textbackslash partial f\textasciicircum2}}
  We prove here Eq.~(\paper41), and thus that $\Lh_\sto$ is convex
  and has only one maximum at $f = \hat f_\sto$ (\cf~\S~\paper3.2.2).

  Deriving Eq.~(\paper38), one obtains
  \begin{align}
    \label{p41-1}
    \frac{\partial^2\ln\Lh_\sto}{\partial f^2}
    &=
    - \frac{1}{f^2*(1-f)^2}*\left(n*[1-f]^2
      + f*[1-f]*\frac{\partial\sum_{i=1}^n \Psto[A_{i, 0} \mid C \cap C']}{
        \partial f}
      + [2*f-1]*\sum_{i=1}^n \Psto[A_{i, 0} \mid C \cap C']\right)
    \notag
    \\
    &=
    - \frac{1}{f^2*(1-f)^2}*\sum_{i=1}^n{\left([1-f]^2
        + f*[1-f]*\frac{\partial \Psto[A_{i, 0} \mid C \cap C']}{\partial f}
        + [2*f-1]* \Psto[A_{i, 0} \mid C \cap C']\right)}.
  \end{align}
  One has
  \begin{align}
    f*(1-f)*\frac{\partial \Psto(A_{i, 0} \mid C \cap C')}{\partial f}
    &=
    f*(1-f)*\frac{-n'*\xi_{i, 0}*(1-f)*n'*\xi_{i, 0}
      +f*\sum_{j=1}^{n'}\xi_{i, j} - (1-f)*n'*\xi_{i, 0}*(-n'*\xi_{i, 0}
      + \sum_{j=1}^{n'} \xi_{i, j})}{\bigl([1-f]*n'*\xi_{i, 0}
      +f*\sum_{j=1}^{n'}\xi_{i, j}\bigr)^2}
    \notag\\
    &=
    -\frac{(1-f)*n'*\xi_{i,0}}{(1-f)*n'*\xi_{i, 0}+f*\sum_{j=1}^{n'}\xi_{i, j}}*\frac{f*\sum_{j=1}^{n'} \xi_{i, j}}{(1-f)*n'*\xi_{i, 0}+f*\sum_{j=1}^{n'}\xi_{i, j}}
    \notag\\
    &=
    -\Psto(A_{i, 0} \mid C \cap C')*\sum_{j=1}^{n'}*
    \Psto(A_{i, j} \mid C \cap C')
    \notag\\
    &=
    -\Psto(A_{i, 0} \mid C \cap C')*\bigl(1-\Psto[A_{i, 0} \mid C \cap C']\bigr),
  \end{align}
  so
  \begin{equation}
    \label{p41-2}
    \begin{split}
      (1-f)^2 + f*(1&-f)*\frac{\partial \Psto(A_{i, 0} \mid C \cap C')}{\partial f}
      + (2*f-1)* \Psto(A_{i, 0} \mid C \cap C')\\
      &= (1-f)^2
      - \Psto(A_{i, 0} \mid C \cap C')*\bigl(1-\Psto[A_{i, 0} \mid C \cap C']\bigr)
      + (2*f-1)*\Psto(A_{i, 0} \mid C \cap C')
      \\
      &= (1-f)^2 - 2*(1-f)*\Psto(A_{i, 0} \mid C \cap C')
      + \Psto(A_{i, 0} \mid C \cap C')^2
      \\
      &= \bigl([1-f] - \Psto[A_{i, 0} \mid C \cap C']\bigr)^2.
    \end{split}
  \end{equation}
  Equation~(\paper41) follows immediately from Eqs.~\eqref{p41-1}
  and~\eqref{p41-2}.
\subsection{Analytic expression of 
  \texorpdfstring{$\hat\partial^2\ln\Lh_\sto/(\hat\partial x_p*\hat\partial x_q)$}{\textbackslash hat\textbackslash partial\textasciicircum2 ln L\_\{s:o\}/(\textbackslash hat\textbackslash partial x\_p \textbackslash hat\textbackslash partial x\_q)}}
  \label{unc_sto}
  Deriving Eq.~(\paper74), one obtains
  \begin{equation}
    \frac{\partial^2\ln\Lh}{\partial x_p*\partial x_q}
    = \frac{1}{\expandafter \Prob(C \mid C')}*
    \frac{\partial^2 \Prob(C \mid C')}{
      \partial x_p*\partial x_q} - \frac{1}{\Prob^2(C \mid C')}
    *\frac{\partial \Prob(C \mid C')}{\partial x_p}*
    \frac{\partial \Prob(C \mid C')}{\partial x_q}\,,
  \end{equation}
  and, because of Eqs.~(\paper74) and~(\paper28),
  \begin{equation}
    \label{der2(Lh_gen)}
    \frac{\hat\partial^2\ln\Lh}{\hat\partial x_p*\hat\partial x_q}
    = \frac{1}{\expandafter\expandafter\hat\Prob(C \mid C')}*
    \frac{\hat\partial^2 \Prob(C \mid C')}{
      \hat\partial x_p*\hat\partial x_q}.
  \end{equation}
  From Eqs.~(\paper33), (\paper74) and~(\paper4),
  one gets
  \begin{equation}
    \label{der2(P_sto)}
    \frac{\partial^2 \Psto(C \mid C')}{\partial x_p*\partial x_q}
    = \sum_{i=1}^n\sum_{j=0}^{n'}{\frac{\partial^2\ln\zeta_{i, j}}{
        \partial x_p*\partial x_q}*\Psto(A_{i, j} \cap C \mid C')}
    + \sum_{i=1}^n\sum_{j=0}^{n'}{\frac{\partial\ln\zeta_{i, j}}{\partial x_p}
      *\frac{\partial \Psto(A_{i, j} \cap C \mid C')}{
        \partial x_q}}.
  \end{equation}
  Using Eqs.~(\paper22) and~(\paper75)
  with $h_k = \sum_{j_k=0}^{n'} \zeta_{k, j_k}$
  and $\Upsilon = \IE[1, n]\setminus\{i\}$,
  the right-most term above becomes
  \begin{align}
    \frac{\partial \Psto(A_{i, j} \cap C \mid C')}{\partial x_q}
    &= \Xi*\frac{\partial\zeta_{i, j}}{\partial x_q}*
    \prod_{\substack{k=1\\ k\neq i}}^n\sum_{j_k=0}^{n'}\zeta_{k, j_k}
    + \Xi*\zeta_{i, j}*\sum_{\substack{\ell=1\\\ell\neq i}}^n{
      \frac{\partial\sum_{j_\ell=0}^{n'}\zeta_{\ell, j_\ell}}{\partial x_q}*
      \prod_{\substack{k=1\\ k\not\in\{i{,}\,\ell\}
        }}^n\sum_{j_k=0}^{n'}\zeta_{k, j_k}}
    \notag \\
    &= \Xi*\frac{\partial\ln\zeta_{i, j}}{\partial x_q}*\zeta_{i, j}*
    \prod_{\substack{k=1\\ k\neq i}}^n\sum_{j_k=0}^{n'}\zeta_{k, j_k}
    + \Xi*\frac{\zeta_{i, j}}{\sum_{j_i=0}^{n'}\zeta_{i, j_i}}*
    \sum_{\substack{\ell=1\\\ell\neq i}}^n\sum_{j_\ell=0}^{n'}
    {\frac{\partial\ln\zeta_{\ell, j_\ell}}{\partial x_q}*
      \zeta_{\ell, j_\ell}*
      \prod_{\substack{k=1\\ k\neq\ell}}^n\sum_{j_k=0}^{n'}\zeta_{k, j_k}}
    \notag \\
    &= \frac{\partial\ln\zeta_{i, j}}{\partial x_q}*
    \Psto(A_{i, j} \cap C \mid C')
    + \Psto(A_{i, j} \mid C \cap C')*
    \sum_{\substack{\ell=1\\\ell\neq i}}^n\sum_{j_\ell=0}^{n'}
    {\frac{\partial\ln\zeta_{\ell, j_\ell}}{\partial x_q}*
      \Psto(A_{\ell, j_\ell} \cap C \mid C')}.
    \label{der(P_sto)_gen}
  \end{align}
  For $\vec x = \skew3\hat{\vec x}_{\sto}$,
  \begin{align}
    \sum_{\substack{\ell=1\\ \ell\neq i}}^n\sum_{j_\ell=0}^{n'}
    {\frac{\hat\partial\ln\zeta_{\ell, j_\ell}}{\hat\partial x_q}*
      \expandafter\hat\Psto(A_{\ell, j_\ell} \cap C \mid C')}
    &= \sum_{\ell=1}^n\sum_{j_\ell=0}^{n'}
    {\frac{\hat\partial\ln\zeta_{\ell, j_\ell}}{\hat\partial x_q}*
      \expandafter\hat\Psto(A_{\ell, j_\ell} \cap C \mid C')}
    - \sum_{j_i=0}^{n'}
    {\frac{\hat\partial\ln\zeta_{i, j_i}}{\hat\partial x_q}*
      \expandafter\hat\Psto(A_{i, j_i} \cap C \mid C')}
    \notag \\
    &= - \sum_{k=0}^{n'}
    {\frac{\hat\partial\ln\zeta_{i, k}}{\hat\partial x_q}*
      \expandafter\hat\Psto(A_{i, k} \cap C \mid C')}
    \label{der(P_sto)_subterm}
  \end{align}
  since the first term on the right-hand side of the first line
  is zero from
  Eqs.~(\paper33) and~(\paper28).
  Finally, combining Eqs.~\eqref{der2(P_sto)}, \eqref{der(P_sto)_gen},
  \eqref{der(P_sto)_subterm}
  and dividing by $\expandafter\hat\Psto(C \mid C')$,
  one obtains
  from Eqs.~\eqref{der2(Lh_gen)} and~(\paper4)
  that
  \begin{equation}
    \label{der2(Lh_sto)}
    \frac{\hat\partial^2\ln\Lhsto}{\hat\partial x_p*\hat\partial x_q}
    = \sum_{i=1}^n{\sum_{j=0}^{n'}{\left(\frac{\hat\partial^2\ln\zeta_{i, j}}{
            \hat\partial x_p*\hat\partial x_q} +
          \frac{\hat\partial\ln\zeta_{i, j}}{\hat\partial x_p}*
          \Biggl[\frac{\hat\partial\ln\zeta_{i, j}}{\hat\partial x_q}
          - \sum_{k=0}^{n'}{\frac{\hat\partial\ln\zeta_{i, k}}{\hat\partial x_q}
            *\expandafter\hat\Psto(A_{i, k} \mid C \cap C')}
          \Biggr]\right)
        *\expandafter\hat\Psto(A_{i, j} \mid C \cap C')}}
    .
  \end{equation}

  For $x_p = x_q = f$, one recovers Eq.~(\paper41) at $f = \hat f_\sto$.
  For other unknown parameters, a numeric computation might be
  simpler\textellipsis{}
  Whatever the method, one can then build the covariance matrix of 
  $\skew3\hat{\vec x}_\sto$
  and derive the uncertainties on the unknown parameters from it
  (cf.~Eq.~(\paper34) and \S~\paper3.2.1).
  % 
  %%%%%%%%%%%%%%%%%%%%%%%%%%%%%%%%%%%%%%%%%%%%%%%%%%%%%%%%%%%%%%%%%%%%%% 

\end{appendices}
% 
%%%%%%%%%%%%%%%%%%%%%%%%%%%%%%%%%%%%%%%%%%%%%%%%%%%%%%%%%%%%%%%%%%%%%%% 
% Spacing between references.
\makeatletter
\def\item{%
\item@orig
}
\makeatother
\addtocontents{toc}{\bigskip\bigskip}
\bigskip\bigskip
\bibliographystyle{perso}
\bibliography{references}
\end{document}